# Atomic beings and the discovery of gravity

G. De Conto[1], G. Franzmann[2*]

November 15, 2015

[1]*Instituto de Física Teorica, UNESP-Universidade Estadual Paulista
R. Dr. Bento T. Ferraz 271, Bl. II, Sao Paulo 01140-070, SP, Brazil,*
[2]*Department of Physics, McGill University, Montréal, QC, H3A 2T8, Canada*

We aim to bring a new perspective about some aspects of the current research in Cosmology. We start with a brief introduction about the main developments of the field in the last century; then we introduce an analogy that shall elucidate the main difficulties that observational sciences involve, which might be part of the issue related to some of the contemporary cosmological problems. The analogy investigates how microscopic beings could ever discover and understand gravitational phenomena.

---
*Electronic address: georgedc@ift.unesp.br; guilherme.franzmann@mail.mcgill.ca

# Contents

# 1 Introduction

Cosmology, the branch of physics concerned with the origin and evolution of the Universe, has been under a revolution over the last century. However, the same revolution that changed completely the way we think about the universe brought new questions that seem far from being settled by the physics community. In order to understand these questions, how they appeared and why they remain to be understood, let us go quickly over the main modern developments of this exciting field.

## 1.1 What a bang!

It all started in 1929, when Edwin Hubble [1] observed that distant galaxies were moving away from Earth. That meant the universe was under expansion instead of being static, which was the common belief by then. The very first astonishing implication of this discovery was that, if the universe was expanding now, it should had been smaller in the past. If this has been always the case arbitrarily to the past (and the universe was also finite in size), thus it meant there was a moment in which the universe was reduced to a point; the whole energy and matter content of it would be concentrated in a point and quantities as energy density, temperature and others would be infinite. It is interesting to note that Alexander Friedmann had already found a theoretical solution to Einstein equations by that time [2] contemplating all those properties. This particular scenario would be known later as The Big Bang.

As you can imagine, even though the universe was indeed expanding, it was not necessarily the case that there was anything like a Big Bang, since we have assumed that it has been expanding arbitrarily in the past. The controversy about Hubble's discovery lasted for almost 40 years, getting to an end due to another discovery: the Cosmic Microwave Background (CMB). In 1964, Arno Penzias and Robert Wilson [3] found out the existence of a radiation in the microwave spectrum (with temperature around $2.7\,K$) permeating the universe, which is understood today as a reminiscent fossil from a very hot ($T \sim 3000\,K$) and dense epoch of the universe in the far past (~ 13.7 billions of years). Because of this discovery, the Big Bang theory became the current paradigm of Cosmology [1].

## 1.2 A cold and dark place

It is interesting to realize that the theoretical understanding coming from Friedmann's solution regarding those discoveries was based on some key assumptions about our universe. Those assumptions are called: *Copernican Principle* and *Cosmological Principle*.

The Copernican Principle basically states that we do not occupy a particularly special place in the universe, that is, whatever conclusion we get here on Earth about the observed data could be applicable to understand other patches of the universe. This remains a principle still, because even though we are travelling in space and time by the movement of Earth, the Solar System, our galaxy, etc, we have not travelled much so far throughout the universe since mankind first appeared on Earth in relation to the relevant cosmological scales[2].

The Cosmological Principle states that the universe is isotropic and homogeneous given some particular scale[3]. In order to understand the reason why there must exist a particular scale for which those properties emerge, we just have to remind ourselves about a bowl filled with water. We know that the liquid is very inhomogeneous and anisotropic at the molecular level, since we can distinguish different molecules, observe their collisions, etc. On the other hand, when we zoom out, all those molecules form together a fluid and their individualities and interactions are no longer relevant to the macroscopic description of the system. Therefore, there is a relevant scale for which the system is observed to be homogeneous and isotropic. Nowadays, we know that this relevant scale for Cosmology is around $100\,Mpc$ [5] according to redshift surveys, where $1\,MpC = 3.08 \times 10^{22}$ meters $\approx 3.2$ million light-years.

Continuing with our little story, it was also in the $70's$ that the idea of *dark matter* was consolidated after the observations of Vera Rubin and Kent Ford [6], due to the observation of rotation of galaxies. The concept in its own had already been introduced in 1932 by Jan Oort and by Fritz Zwicky [7] in 1933, though in both cases the evidences were not robust.

---

[1] It is important to note that initially the Big Bang was understood literally as the beginning of the universe. Today we refer to Standard Big Bang Cosmology, which means we acknowledge there was a moment that the universe was very hot, dense and much smaller than now, without any necessary implications about its origin.

[2] The galaxy velocity is around 2 million $km/h$ [4], that means we have travelled approximately 0.35 light-years since humans appeared on Earth, about $200\,000$ years ago. The cosmological scale is around 3.2 millions light-years, as we will see.

[3] We recall that homogeneity implies that the relevant system is undistinguishable after spatial translations are considered, while isotropy means that it remains the same once observed from different directions.

The problem that motivated the introduction of this new idea was the following: comparing direct observation of visible/luminous matter (the stuff that we are made of) with observations of gravitational effects in the CMB and gravitational lensing, Vera Rubin found a huge difference between the visible matter observed and the amount of matter necessary to produce those effects. Thus, the data indicated that there should be much more matter in the universe than what we could observe directly. This invisible matter was called dark matter, which we now know accounts for around 85% of the matter in the universe.

The last piece of the Cosmological scenario was settled in 1998: the *dark energy*. This piece is an enigma by its own since we still have basically no clue of what it is, even though we have several proposals [4]. As we saw above, it was known that the universe was under expansion since the 30's. However, only in 1998 and 1999 we discovered that this expansion was accelerated, against all the odds, by independent observations of Saul Perlmutter's and Brian Schmidt, Adam Riess' teams [9] [5]. Even more interesting is the fact that the current observational data implies that the amount of dark energy necessary to provide this accelerated expansion must correspond to about 68 % of the energy content of the universe [10].

Looking at the theoretical side as well, everything we just described was only possible due to the comprehension provided by the gravitational theory of Albert Einstein, formulated in 1915 [11] and known as *General Relativity*. The most relevant feature of this theory is the understanding that matter and energy deforms the spacetime structure, which is the "fabric" of the universe.

## 1.3 The big picture

In the end of the day, our mosaic is formed of several parts: in the experimental side, the recognition that our universe is made of 68% of dark energy, 27% of dark matter and 5% of visible matter [10]; in the theoretical side, the General Relativity theory. Those two sides are glued together in a framework called $\Lambda CDM$ model, where $\Lambda$ is associated with dark energy, while CDM stands for *cold dark matter*[6]. This model contains only six free parameters, *i.e.*, quantities that are not predicted by the theory and are free to be adjusted according to the observed data. Among them, there are the age of the universe and the densities of the dark and visible matter. This is incredible! Just to bring some perspective, in the realm of the microscopic world the Standard Model of Particle Physics needs 19 free parameters [12].

---

[4] The first of them, known as cosmological constant, was proposed by Albert Einstein in 1917 [8]. Curiously, it was recognized by him as his biggest mistake in life; by the end of the last century this mistake was shown to be much closer to the truth than expected at Einstein's time.

[5] Naively, one would expect that the universe would be expanding in a decelerated fashion as the worst case scenario because the only relevant force on Cosmological scales, gravity, is attractive.

[6] *Cold* means that it moves slowly in relation to light speed and *dark* because it does not interact with electromagnetic radiation.

There are no doubts that knowing a description for a system of the size of the universe (literally!) with only six free parameters is impressive. Despite that, when we recall that we really understand just 5% of the content of this system, around 27% is partially understood and 68% remains unknown, we need to continue our intellectual incursions so that we achieve an even better description of the universe.

Hence, motivated by the obscurities that still remain in Cosmology, we introduce the following analogy that shall be appreciated differently by different readers: *i*) for the lay public and scientists from other fields, to elucidate the scientific difficulties of observational science as opposed to highly controlled lab experiments; *ii*) for the cosmologists colleagues, to bring a new provocative perspective of how we have been developing our field of knowledge and incite questions about it; *iii*) for the authors, a flight on the wings of imagination.

## 2   How would atomic beings see the world?

Let us imagine a fictitious universe where we aren't the only intelligent beings; a universe also inhabited by creatures much smaller than us that like to study physics as well. Such microscopic beings live in distance scales of the order of the atomic radius, about $1\,\text{Å} = 10^{-10}\,m$, and time scales on the order of $10^{-16}\,s$, the time needed for an electron to go around a hydrogen atom in the Bohr approximation. From now on, whenever we say *we*, we are talking about the human beings living in a macroscopic world, and whenever we say *them*, we refer to the atomic beings. We represent our little friends in the figure below:

Figure 1: Yes, they are quite small!

From the time and space scales involved, it is easy to imagine that the microscopic beings have a world view different than ours. For them, quantum mechanics and quantum field theory (or some version of these theories) are the tools to describe everyday phenomena. Our classical physics, however, is something that they are unaware of, at least in a first glance.

## 2.1 Seeing structures

Since these microscopic beings live in time scales much smaller than ours, the world for them goes in slow motion compared to our view. So, when they encounter an atomic nucleus, what they would see is a bunch of protons and neutrons in random positions. At first, it is hard to make sense of that agglomerate of particles. Looking once, they are in a given set of random positions, looking again later they are in another set of such random positions, so, how to make sense of that? Being smart creatures, they decide to observe and record these positions on several time instants, covering a long stretch of time (long for them, but just a blink of an eye for us). After doing so, the microscopic beings realize that even though on each time instant the protons and neutrons are just scattered in space, in long periods of time these particles occupy always the same region of space. This is their first step to recognize structure in nature, and that is how they discover the existence of the *nucleus*.

But, nucleus of what? Taking a broader look on the spatial scale, they see a different kind of particle, the electron. Such electrons are far from the just found nucleus, they as well look randomly scattered in space. Once again, they decide to record the electronic positions over a long period. Analyzing their data afterwards they see that the electrons travel[7] haphazardly throughout space, however, they notice that some electrons usually stay around a given region of space, while another group of electrons stick to a different region, and so on. Now they find another level of structure, they realize that the electrons move around the nucleus, and that different groups of electrons occupy distinct regions, making what we call *electronic shells*. This big structure, involving the nucleus and the electrons around it, is our familiar *atom*.

Once the atom is recognized as a structure, the atomic beings are now familiar with their most natural spatial scale, allowing them to move forward in the discovery of the world. Through careful observations they realize different size conglomeration of atoms, moving so slowly to them that this movement is barely noticeable. Those awkward objects can be very simple or much complicated, even combining different atoms in themselves. In the end, our microscopic friends decide to call those arrangements *molecules*. But that was not the end of the story! Intrigued as they could about such large structures, different myths were created by some trying to fill their lack of knowledge, while others kept recording data. It is no surprise that the second group, after putting a huge amount of time and effort to make sense of what they observed, discovered that the molecules first seen as odd and sparsely distributed, actually made up together much larger compositions, which we call *condensed matter*. In other words, the solids and liquids phases of composite matter.

We, human beings, follow a similar approach to understand structures that go beyond our everyday space and time scales. When we look to the night sky, all we see are bright spots scattered all over the space. However, watching the sky for long enough, we start to realize that most of these spots move together, as if fixed in the sky (the stars), while others seem roaming alone (the planets). From such observations we can build a picture of what is going on in the cosmological scale. We notice planets going around the stars (specially in the Solar System) forming planetary systems, we also notice stars clumping together making galaxies. From these kind of observations of the atomic world, the small creatures have an experimental picture of the wave functions from quantum mechanics. However, they still don't know why these structures exist, they just know they do.

---

[7] Obviously, we are using daily intuition here. Actually, according to Quantum Mechanics the atomic beings would keep track of the probability of an electron appearing in some place, then in another, and so on.

## 2.2 Why structures?

Now, in an effort to understand what is binding all those particles together, the atomic creatures decide to take a particle physicist approach, observing the behavior of pairs of particles and how they interact with each other. For their new studies, they move to an isolated part of space, where there are only two electrons. It doesn't take long for them to realize that electrons repel each other. Moving on with their journey, they end up inside a candle's flame, a hectic environment full of ions and free electrons moving around and emitting light. Fortunately for our atomic scientists, it all happens in a slow pace, where they can count the number of electrons, protons and neutrons on the ions. Thus, they realize that whenever two ions have more protons than electrons, they repel each other, and if one has more electrons than protons, they attract each other. Conclusion, electrons and protons have opposite charges. However, the neutrons influence these attractions and repulsions too, as if they create some sort of resistance to movement on the ions, either attracting or repelling each other. That is a hint to a property they haven't thought before, the neutrons have no electric charge, but they have *mass*. This is corroborated when they think about the several different ions, that isn't just the difference in electric charge that affects their movement, but also the number of particles in it. Putting all these effects together the atomic creatures figure out *electromagnetism* and *mechanics* as well.

But, if the protons have the same charge, how come they stay together to form a nucleus? Not to mention that the neutrons in there have no electric charge, therefore, something is missing from the picture, some other force must be present. Well, the microscopic creatures have the scenario needed to propose (and study) a new interaction, what we call the *strong nuclear force*. The protons that they have seen before were repelling each other, just as the electromagnetic interaction predicts, however, when in the nuclei, they stick together. The conclusion: the always attractive strong nuclear force has a short range, becoming feeble for distances greater than the radius of a nucleus, but stronger than the electromagnetic interaction at short distances. For distances greater than the nucleus, the electromagnetic force dominates, and the protons push each other away.

To be sure that their theories are correct, the atomic beings keep on studying several nuclei, from different atoms. But, when observing some nuclei, they encounter something new. Sometimes a neutron turns into a proton and emits an electron in the process[8], something very strange and unaccounted by the forces they have found so far. Further inspection of these phenomena leads them to the discovery of a new interaction, that we call *weak nuclear force*[9].

---

[8] Besides emitting an electron, the neutron also emits an anti-neutrino, but let us ignore it just for the sake of the argument.

It is no surprise that the electromagnetic interaction was the first known by those beings. As we live in the macroscopic realm, having the most natural time and spatial scales being the meter and seconds, the gravitational interaction rules our world and it was the first one described; it passed almost two centuries before a proper description of electromagnetism was also obtained. Therefore, a much similar process happens for the atomic creatures, because we know their scale is ruled predominantly by the electromagnetic interaction.

So far, our microscopic creatures have found three forces: electromagnetic, strong and weak (not to mention the influence of mass on the trajectories). But, as we all know, there is still one more to go, an interaction so weak in the microscopic scale that it is going to pose a big challenge for them.

## 2.3 What about gravity?

Living in such small scales provides several advantages when one attempts to understand the microscopic world. As we have discussed in the last section, electromagnetism is the "daily" interaction for the atomic creatures and it is only a matter of looking at scales $10^5$ times smaller than the atomic size to have a direct look at the atomic nucleus. Thus, the subatomic world for them is as far as the bacterias are for us, which can be easily observed with an optical microscope. On the other hand, the study of the large becomes much more difficult. We have seen their journey to understand the existence of molecules, and even condensed matter. The same natural scale that allowed them a rapid understanding of the microscopic world now constrains every possible experiment they can make to keep their exploration into larger scales. But, there is still hope for our little friends!

Being good scientists as they are, at some point they figure out how to make observations up to the meter scale, which is $10^{10}$ times their own. Within such incredible range, a whole new world reveals to them. Among all the new unimaginable structures they now see, a peculiar reddish one calls their attention. Oddly enough, this object seems to stand still and isolated in space. This humongous red shape, far, and almost static for the atomic beings, is nothing more than an ordinary apple falling towards the ground for us. Being in the meter scale, the apple not only is of the order $10^{10}$ times bigger than our friends, but also the time scale that it takes to fall to the ground is about $10^{16}$ times longer than their time scale - what seems an infinite amount of time![10]

Now, things start to become really difficult. Constrained by their size, there is no way our friends could make any experiment to investigate what is going on on the large scales. Different from the initial discoveries they had, in which they were able to play around with electrons, protons, neutrons and other subatomic particles, only observational science is available for them this time. Needless to say, this will be the seed for confusion as their journey continues.

---

[9] The weak nuclear force and the electromagnetic force are correlated in their origins, making the electroweak force. However, for our analogy, it is better to treat them separately.

[10] Just to put things in perspective, something that is $10^{10}$ meters big is approximately $10^{-5}$ light-years, of the same order as the distance from Earth to the Sun, while a phenomenon that takes $10^{16}$ seconds is around 320 millions years long.

Well, having so far understood three interactions and probably being really proud of the models already established, their first attempt to comprehend the new discoveries would be made using what is already known. However, they know that the strong and weak nuclear interactions are short range and should not have a direct influence in the newly observed phenomena. Therefore, it should be the infinite range interaction known to them, electromagnetism, that should account for the large structure data.

Wait a second! - some of us could think. We know that the fall of an apple could not be possibly explained by electromagnetism. In fact, we have another interaction that accounts for this, *gravity*. Although this is true, *a priori* there is no reason to believe our little friends would take gravity for granted. As we just said, once they have a simple and elegant framework made up of the three interactions known to them, our history of science tells us that they would stick to that and do whatever it takes to make sense of the world with the tools already developed.[11] Maybe, before judging them, let us have a look on what they could do with what they already know.

### 2.3.1 An odd idea

After the understanding of molecules, chemical bounds and condensed matter physics together with the idea of inertia/mass, the atomic creatures would be able to write down the following equation for the electric interaction between bodies 1 and 2,

$$a_2 = k_e \frac{q_1}{r_{12}^2} \frac{q_2}{m_2}, \tag{1}$$

where $m_2$ is the mass of some falling object, $a_2$ is its acceleration, $r_{12}$ is the distance between the two objects that are interacting, $q_i$ are the electric charges of those objects and $k_e \approx 10^9 \, \text{N}\,\text{m}^2\,\text{C}^{-2}$ is an effective macroscopic coupling of the electrical counterpart of the electromagnetic interaction. We can expect them to have developed this macroscopic description for electromagnetism once they have studied larger systems, such as molecules and ions. This is the kind of description fit to the case where they studied the ions in a flame, as discussed previously.

Now, when the microscopic creatures look to their surroundings, everything seems to be electrically neutral in average. However, due to their size, they might be mistaken about it when larger scales start to be considered; after all, it would be too hard for them to keep control of every single charged particle floating around them. So, it is no surprise that they could postulate those gigantic objects were charged. Actually, they would go beyond. Realizing that large structures have more and more particles and also reminding that particles have mass, it would be reasonable to assume that the net charge of an object would be proportional to its volume and, consequently, to its mass. Thus,

---

[11] This happened several times in the history of physics and many of them misled us. Two examples are Ptolemy epicycles and the aether theory.

in the end their assumption could be translated as

$$q_i = \alpha m_i, \tag{2}$$

where $\alpha$ is a constant and certainly much less than 1, otherwise they would notice a net charge in their own scale. For the sake of the analogy, let us assume that their experiments guarantees that $\alpha \lesssim 10^{-10}\, C\, Kg^{-1}$.

Next, the creatures with the observations of different huge objects moving in space (actually, those objects are all falling down to the ground; but they can barely see an apple, so imagine about understanding the notion of ground!), they are able to conclude that all of them have the same acceleration, around $a \approx 10\, m/s^2$! That is astonishing for them! Now, they can write down

$$\begin{aligned} a &\sim k_e \frac{q_1}{r_{12}^2} \alpha \\ \frac{q_1}{r_{12}^2} &\sim \frac{a}{k_e \alpha}, \end{aligned} \tag{3}$$

which is very similar to Newtonian gravity close to Earth's surface, where we can write

$$\frac{M_1}{r_{12}^2} \sim \frac{g}{G}, \tag{4}$$

being $M_1$ the mass of the Earth, $G$ the Newton's constant and $g$ the constant gravitational acceleration. We know that in this case we can write $r_{12} \approx R$, where $R$ is the radius of the Earth. This approximation is only valid for bodies close to the surface of the Earth, because $r_{12} = z + R$ (where $z$ is the height above the surface) and $R \gg z$, which guarantees a constant acceleration for all of them.

Having the atomic beings concluded that those gigantic objects have approximately the same acceleration, they could similarly write down

$$\begin{aligned} \frac{q_1}{R^2} &= \frac{a}{k_e \alpha} \\ \frac{q_1}{R^2} &= 10^2 \frac{C}{m^2}, \end{aligned} \tag{5}$$

where for them it would be not clear what $R$ represents: it could be the radius of a large object or just the distance of the falling objects to the object's center of charge. Independently of that, if their model is correct, there is an unexpected conclusion: there should be another object/structure with opposite net charge at some place that would be attracting all those falling objects and its charge should be enormous by equation (5)[12]. Not really seeing this new structure so far, they decide to call it *the big dark charge*[13].

---

[12] It is curious that having assumed $\alpha \sim 10^{-10}\, C\, Kg$, one could insert in the radius of the Earth as $R$ in (5) and would encounter a mass for the Earth as being of the order $10^{25}\, Kg$, off by one order of magnitude!

[13] Our small friends are, actually, discovering the Earth in their own model, though the properties of our planet are a mess! Of course, instead of the mass of the planet playing the role of the charge through a gravitational interaction, they supposed it carried an electric charge to describe falling objects through the electric force.

We see that the atomic creatures are really smart and doing an amazing progress in their understanding of the world. One could bet that only an Einstein-like creature would be able to propose such strange description of the large scale world having lived his or her full life being so tiny!

### 2.3.2 Shouldn't apples repel themselves?

Of course not! Oops, that is *our* answer to the question.

The model so far proposed by our small friends implies that every single falling object should have the same charge, since that would be the only way they would be attracted to the big dark charge; in another words, that is how things fall.

Because they have the same charge, apples repel other apples. Therefore, the model should be ruled out, right? Well, they can investigate that. The acceleration between two apples is given by

$$a_{ap,2} = k_e \frac{q_{ap,1}}{r_{12}^2} \frac{q_{ap,2}}{m_{ap,2}}, \tag{6}$$

where "*ap*" stands for apple. Having observed many different falling apples in the space, they realize that they can be found as close as a tenth of a meter, that means, $r_{12}^2 \sim 10^{-2}\, m$. Also, from their understanding so far, the creatures would be able to estimate the mass of an apple looking to its size. Of course, they would have to make some assumptions about density, for instance, but that would not change much the order of magnitude of the results. Thus, let's say that a typical apple has a mass around $0.1\, Kg$. Hence, putting in all the numbers,

$$a_{ap,2} \sim 10^{-10}\, m/s^2. \tag{7}$$

This is $10^{11}$ times smaller than the acceleration between an apple and the big dark charge, being completely negligible for their observations. In fact, in their units, $a_{ap,2} \sim 10^{-32}\, Å/(Bohr's\, time)^2$.

At the end of the day, it seems that our little friends were able, after all, to explain the falling objects phenomenon. While new observations with different results are not obtained by them, there will be no reason to expect the conservative part of the atomic scientific physics community to put in check their electromagnetic model. Unfortunately, they are fundamentally completely mistaken.

### 2.3.3 Getting it right

The atomic beings seem to be lost for the moment, but their own journey is not finished yet. There are various ways out of their misunderstanding concerning gravity and we can, more or less, predict what could happen in their path from now on.

One possibility is observing a gravity-induced quantum interference. This is a known experiment in which we can directly see the influence of the gravitational force in the quantum realm.

The setup is the following: consider a monoenergetic beam of particles and split it in two, so that each part takes a different path, as shown in Fig. 2. Note that $l_2$ is a vertical distance, each path has its horizontal part at a different height and, therefore, under a different gravitational potential.

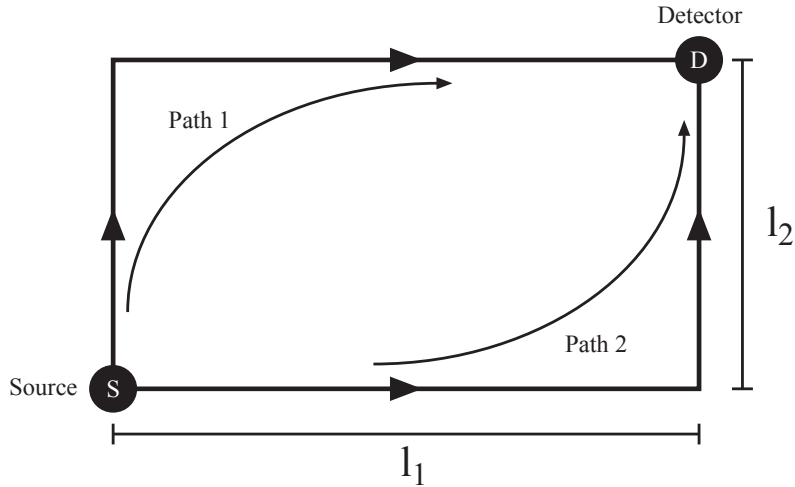

Figure 2: The apparatus necessary to the experiment. The point $S$ indicates the source of the beam, $D$ represents the detector, $l_1$ is the horizontal separation between them, and $l_2$ is the vertical separation.

A very easy way to understand what happens can be found in [13]. In summary, since the potential energy is time-independent, the total energy is constant,

$$\frac{\mathbf{p}^2}{2m} + mgz = E, \qquad (8)$$

where $\mathbf{p}$ is the particle momentum, $m$ is its mass, $z$ is height above ground, $g$ is the gravitational constant, and $E$ is its total energy. A difference in height of each beam implies a difference in $\mathbf{p}$, or $\lambda$, the wavelength of the particles, according to De Broglie formula which relates the matter wavelength to its momentum by $\lambda = h/p$. Therefore, an interference pattern is observed. In fact, the phase difference between the two paths is given by

$$\phi_1 - \phi_2 = -2\pi \frac{m^2 g l_1 l_2 \lambda}{h^2}, \qquad (9)$$

where $m$ is the mass of the particles in the beam, $h$ is the Planck constant, $g$ is the gravitational acceleration close to Earth's surface and $l_1$ and $l_2$ are the lengths indicated in the Fig. 2.

Therefore, the small creatures could in principle detect this phase difference and start wondering about what causes this effect. At some point, this could lead them to the correct proposition of a new interaction in their Standard Model of physics. There is only one drawback here: in order to this effect to be relevant, either the time involved or the apparatus size should be huge for them. For a more detailed discussion, see appendix A.

Another observation that they could do is related to their understanding of electromagnetic theory. Once it is known that any charged particle should radiate when accelerated, this could be a check for their model. The only problem here is that those supposed charged falling apples in this model would be emitting a negligible amount of radiation anyway, since their acceleration is so small. So, it is a matter of experimental resolution to conclude that the falling apples are not radiating anything and, therefore, that they do not have a net electric charge.

Although there is a whole framework called gravitoelectromagnetism [14] which makes use of similarities between electromagnetic theory and gravity, there are also many differences between both interactions. Certainly, many experiments could be proposed that would have inconsistent outcomes if one only assumes one of those two interactions. However, knowing that these experiments in general involve either a scale much larger than the atomic one, making our friends mere observers, or a tremendous experimental resolution, most of them would be pretty hard to be realized by the atomic creatures, cursing them to remain mistaken by an undetermined period of time. Until either one of the experiments is feasible or some genius-like creature is able to reformulate the whole understanding they have.

## 3 What about us?

In relation to cosmological scales, we are just like the atomic beings regarding the meter/second scale. Not only are we constrained by our own evolutionary outcome, but also by our physical scales of space, time and energy. Now, it is time to go back to our own world.

As the atomic beings, our story is far from the end. In the introduction, we recalled that our Standard Cosmology Model has only six free parameters, but we currently are relatively ignorant about 95% of the matter-energy content of the universe. Analogous to their story, Cosmology is accessible to us only through observations for now and the scales involved are enormously larger than ours, making it very hard to have a proper intuition of the cosmological phenomena and accumulate sufficient data to exclude different models. For example, if we were able to observe the universe from now on to millions of years ahead, we could know about whether dark energy is just a cosmological constant or not.

But, are we in the same situation as the atomic beings, in which we are unaware of other interactions? In this case we have some options, models with extra dimensions, as String Theory, can give us new phenomena, arising from the effects of the known interactions extending into these unseen dimensions. From the point of view of particle physics, models that extend/change the gauge symmetries pose as candidates (supersymmetry, 3-3-1 models, Left-Right models, etc.). These types of extensions bring new particles and their interactions into the scene, which may bring changes to Cosmology and particle physics alike.

On the other hand, if we know all the interactions that exist (strong, electroweak and gravitational), we are faced with a different challenge: reviewing all our current models. Let's take dark energy as an example. We know that the universe is under accelerated expansion based on what we know about gravity, but if Einstein's General Relativity is incomplete, it may be the case that dark energy is something quite different than just a cosmological constant as in the Standard Cosmology Model. Ideas like modified gravity on very large scales, quintessence, brane cosmology, etc, modify what we already know about gravity and also lead to new effects, that may mimic the effects of a mere cosmological constant in our current theory of gravitation. Since we have a limited experimental picture of the Cosmos, we must consider the possibility of phenomena that force us to change our theories just like the atomic beings. Dark matter itself is one of those phenomena, which lead particle physicists into proposing new models to account for it. The *status quo* is that Einstein is right, and that we must accommodate dark energy and dark matter into our picture of Nature. However, correcting Einstein's General relativity is a possibility that cannot just be discarded, even though it may seem the least likely case for those scales.

Either way, we are living in a cosmological conundrum, where our theories don't fit each other perfectly (for instance, we cannot match gravity and quantum mechanics in a single framework) and our experimental knowledge is limited. Maybe we are wrong in both ways: we are unaware of new interactions, and the ones we know are incorrectly described by us (sort of a "worst case scenario"). This puts us in a situation where we have always been, meaning that we have to explore further and deeper into Nature, that we have to improve our models, making them ever more accurate and cohesive. Maybe one of the alternatives we already have proves to be the right answer, maybe we need a new picture; whichever way we go, there is work to be done.

No matter what, as for the atomic beings, there is always the way out in which we have direct access to the relevant scales, that is, being able to do Cosmological experiments and then to rule out theories in a similar way as we do today in Particle Physics. Unfortunately, this is a science fiction scenario for now, as in the story "The Last Question", by Isaac Asimov [15].

Finally, we hope that this work is a reality check, a ludic reminder about the current research that is being done in Cosmology, which despite its successes, it is rather limited compared to other fields of physics.

## 4 Acknowledgments

We would like to thank Renato da Costa Santos, Elisa Gouvea Maurício Ferreira, and Maurício Girardi Schappo for valuable discussions and proofreading and all the friends that gave us suggestions to the manuscript. GDC is supported by CNPq and GF is supported by CNPq through the Science without Borders (SwB).

## 5 Appendix A

Plugging in some numbers for a neutron, we have

$$\begin{aligned}\phi_1 - \phi_2 &= -2\pi \frac{m^2 g l_1 l_2 \lambda}{h^2} \\ &\sim 4 \times 10^{14} l_1 l_2 \lambda.\end{aligned} \qquad (10)$$

On the other hand, the velocity of a neutron is given by,

$$\begin{aligned}v &= \frac{c}{\sqrt{1 + \frac{\lambda^2 m^2 c^2}{h^2}}} \\ v &\sim 10^{-15} \frac{c}{\lambda}.\end{aligned} \qquad (11)$$

Therefore, the total time for the beam to go over its path is,

$$\begin{aligned}\Delta t &= \frac{l_1 + l_2}{v} \\ &\sim 2.5 \frac{l_1 + l_2}{l_1 l_2} \frac{\Delta \phi}{c}.\end{aligned} \qquad (12)$$

The most general situation for the apparatus is to assume that $l_1$ and $l_2$ represent different length scales, so let us call them $L$ and $l$, $L \gg l$. Thus,

$$\begin{aligned}\Delta t &\sim 2.5 \frac{\delta \phi}{lc} \\ &\sim 10^{-8} \frac{\delta \phi}{l_{meter}} s \\ \Delta t &\sim 10^{8} \frac{\delta \phi}{l_{meter}} (Bohr\ time),\end{aligned} \qquad (13)$$

where $l_{meter}$ means the size of the interferometer arm given in meters. Therefore, if an atomic scientist aims to measure a relevant phase difference, that either means having an incredibly long measurement or a gigantic apparatus in their units. Much similar with our situation concerning some cosmological phenomena. This is an explicit example of how scales can constrain experiments.